\documentclass[a4paper,fleqn,usenatbib]{mnras}

\usepackage{graphicx}
\usepackage[percent]{overpic}
\usepackage{lipsum}
\usepackage{multirow}
\usepackage{color}
\usepackage{rotating}
\usepackage{mwe,tikz}
\usepackage[percent]{overpic}
\usepackage{mathtools}
\usepackage{physics}
\usepackage{amsmath}

\newcommand{\msun}{${\rm M_{\sun}}$}

\def\ltsima{$\; \buildrel < \over \sim \;$}
\def\simlt{\lower.5ex\hbox{\ltsima}}
\def\gtsima{$\; \buildrel > \over \sim \;$}
\def\simgt{\lower.5ex\hbox{\gtsima}}
\def\kms{{\rm\,km\,s^{-1}}}
\def\pc{{\rm\,pc}}
\def\kpc{{\rm\,kpc}}

\def\msun{{\rm\,M_\odot}}


\def\deg{^\circ}

\def\Gyr{{\rm\,Gyr}}

\def\masyr{{\rm\,mas \, yr^{-1}}}

\def\ltsima{$\; \buildrel < \over \sim \;$}
\def\gtsima{$\; \buildrel > \over \sim \;$}

\def\Khyati#1{\noindent{{\color{blue} \bf[$\diamondsuit$ #1]}}}

\title[The Solar Velocity Vector]{Measuring the Sun's Motion with Stellar Streams}

\author[Malhan \& Ibata]{
Khyati Malhan,$^{1}$\thanks{E-mail: khyati.malhan@astro.unistra.fr}
Rodrigo A. Ibata,$^{1}$
\\
$^{1}$Universit\'e de Strasbourg, CNRS, Observatoire astronomique de Strasbourg, UMR 7550, F-67000 Strasbourg, France\\
}

\date{Accepted 2017 June 23. Received 2017 June 23; in original form 2017 March 13}

\pubyear{2017}

\begin{document}
\label{firstpage}
\pagerange{\pageref{firstpage}--\pageref{lastpage}}
\maketitle

\begin{abstract}
We present a method for measuring the Sun's motion using the proper motions of Galactic halo star streams. The method relies on the fact that the motion of the stars perpendicular to a stream from a low-mass progenitor is close to zero when viewed from a non-rotating frame at rest with respect to the Galaxy, and that the deviation from zero is due to the reflex motion of the observer. The procedure we implement here has the advantage of being independent of the Galactic mass distribution. We run a suite of simulations to test the algorithm we have developed, and find that we can recover the input Solar motion to good accuracy with data of the quality that will soon become available from the ESA/Gaia mission.
\end{abstract}

\begin{keywords}
Sun: fundamental parameters - Galaxy: fundamental parameters - Galaxy: structure - stars: kinematics and dynamics - Galaxy: halo
\end{keywords}

%%%%%%%%%%%%%%%%%%%%%%%%%%%%%%%%%%%%%%%%%%%%%%%%%%

%%%%%%%%%%%%%%%%% BODY OF PAPER %%%%%%%%%%%%%%%%%%

\section{Introduction}
In physics, it is always of fundamental importance to know the properties of the frame from which measurements are being made. Observations of astrophysical or cosmological systems, using satellites or ground based telescopes, can  be corrected into the Heliocentric frame with ease. However, knowledge of the Sun's Galactic velocity $\bmath{\textbf{V}}_{\odot}$ is required to transform any observed Heliocentric velocity into the Galactic frame. This is necessary, for instance, for scientific interpretation when studying Galactic dynamics or for correcting the motion of many extragalactic systems (see, e.g., \citealt{2016MNRAS.456.4432S}). Moreover, the related circular velocity at the Solar radius ($v_{circ\odot} \equiv v_{circ}(R_{\odot})$) also serves as a crucial constraint on the mass models of the Milky Way (e.g., \citealt{Dehnen1998Massmodel}). Therefore, the determination of $\bmath{\textbf{V}}_{\odot}$ is a crucial task of Galactic astronomy.

It is important to realise that the Sun's Galactic velocity $\bmath{\textbf{V}}_{\odot}$ needs to be measured with respect to some other reference or tracer. A conceptually straightforward way to measure $\bmath{\textbf{V}}_{\odot}$ is to determine the Sun's motion with respect to a presumed motionless object with respect to the Galaxy. Such measurements are derived from the observed proper motion of Sgr $A^{\ast}$ \citep{Reid2004}. But such an approach requires an accurate measurement of $R_{\odot}$ and a critical assessment of measurements coming from the dense region at the Galactic centre, with its complex dynamical mix of gas, dust, stars and central black hole. 

Alternatively, analyses can be based on local tracers, where one assumes that the Solar motion can be decomposed into a circular motion of the Local Standard of Rest (LSR) plus the so-called peculiar motion of the Sun with respect to the LSR: $\bmath{\textbf{V}}_{\odot} = \bmath{\textbf{V}}_{circ\odot} + \textbf{V}_{p \odot}$. Such a study was presented by \citet{Dehnen1998SunHipparcos}, who applied the Str\"omberg relation (\citealt{Stromberg1946}) in their method to a sample of $\sim$ 15 000 main-sequence stars from the Hipparcos catalogue. They determined the peculiar velocity to be $\textbf{V}_{p \odot}= (10.0\pm0.36, 5.25\pm0.62, 7.17\pm0.38)\kms$ (in the conventional $U,V,W$ directions, respectively). However, \citet{Binney2010} caution against this employment of Str\"omberg's Relation and illustrate, using their chemodynamical model of the Galaxy, that the metallicity gradient of the disk population causes a systematic shift in the estimation of the kinematics of the Sun. They describe an alternative method to determine the Sun's velocity with respect to the LSR from the velocity offset that optimizes their  model  fit  to  the  observed  velocity  distribution.  Using  their chemodynamical  evolution model of the  Galaxy, described in \citet{Binney2009a}, they find the Sun's peculiar motion to be $\textbf{V}_{p \odot}=$ ($11.1^{+0.69}_{−0.75}$, $12.24^{+0.47}_{−0.47}$, $7.25^{+0.37}_{−0 .36} \kms$) and estimate roughly the systematic uncertainties as $(1.0, 2.0, 0.5) \kms$. However, their approach has the disadvantage being based on an extensive modelling of the Milky Way, and hence of being sensitive to the adopted approximations in dynamics and chemistry.

Once the Sun's peculiar velocity is known, one still needs to add the velocity of the LSR to obtain the Sun's velocity with respect to the Galaxy. It is interesting in this context to examine what it is currently possible to measure with respect to nearby tracers. In a recent contribution, \citet{Bobylev2013} determined the Solar Galactocentric distance $R_{\odot}$ and Galactic rotational velocity $v_{circ\odot}$, as modified by \citet{Sofue2011_to_support_bobylev}, using data of star-forming regions and young Cepheids near the Solar circle. Based on a sample of 14 long-period Cepheids with Hipparcos proper motions they obtained $R_{\odot}=7.66\pm 0.36\kpc$ and $v_{circ\odot}=267\pm17\kms$. However, with a sample of 18 Cepheids with UCAC4 proper motions (among which 2 were taken from Hipparcos) they found $R_{\odot}=7.64\pm 0.32\kpc$ and $v_{circ\odot}=217\pm 11 \kms$. The difference in the derived $v_{circ\odot}$ values highlights the difficulty of such measurements, and their sensitivity to the adopted tracers and data. Masers located in regions of massive star-formation have also yielded estimates of the LSR motion ($254\pm16\kms$, \citealt{Reid2009}; $236\pm11\kms$, \citealt{Bovy2009}), though these results are derived from a small number of sources (18) and require knowledge of the velocity lag of the masers with respect to circular motions \citep{Rygl2010}.

Here we will examine the power that streams hold to constrain the Solar velocity with respect to the Galaxy. A growing number of stellar streams have been detected in recent years from the Sloan Digital Sky Survey and Pan-STARRS (\citealt{Odenkirchen2001}; \citealt{Grillmair2006}; \citealt{Bernard2014panstarstream}; \citealt{2016MNRAS.463.1759B}). The most recent contributions come from the ATLAS survey and surveys with  CTIO/DECam (the Atlas stream in \citealt{Koposov_ATLAS2014}, and the Eridanus and Palomar 15 streams in \citealt{DES_Streams2017}). The kinematics of these structures will soon be revealed in the second data release of the Gaia mission survey (\citealt{GaiaDR12016}). Gaia would possibly also uncover many new low-contrast star streams that are currently below detection limits in star-count surveys.

The key insight about streams that we exploit here is that stream stars move approximately along their orbits, not perpendicular to them. That is, the velocity vector of a stream star in the Galaxy's rest frame must be a tangent vector to the orbit of this extended structure at that stellar position. Thus, if we measure any motion perpendicular to the orbital path of the stream at this position, we must reconcile it with the apparent (reflex) motion that emerges due to the motion of the observer's frame (from the Sun). Hence, by measuring this perpendicular motion vector for the stars in the streams, we can constrain the Sun's velocity in the Galaxy. This is not an entirely new insight. Several studies made in the past that have examined the kinematics of stellar streams have had to include (implicitly or explicitly) the Solar motion or the circular velocity of the LSR as a nuisance parameter to fit the stream in kinematics space \citep{Ibata2001, Koposov2010, 2015ApJ...803...80K, 2016arXiv160901298B}. Although, \citet{Dehnen2004thinorbit} comment on the fact that in their modelling of the tidal dissolution of the Palomar 5 globular cluster, the ensuing stream actually deviates slightly from the path of the orbit (rather than not at all, as would be naively expected). Such an offset of the stream structure from the underlying orbit could in principle create a bias in the Solar velocity measurements using the method proposed in this contribution. However, we show that by analysing multiple streams on different orbits the bias is largely elliminated.

A similar, but less general version of the idea presented here, was explored in \citet{Majewski2006}. They suggest measuring the Sun's reflex motion using the Sagittarius (Sgr) stream (\citealt{Ibata2001}; \citealt{Majewski2003SagStream}), making use of the fact that the orbital plane of the Sgr stream is polar and that the Sun lies close to this plane. Thus the $V$ motions of the stars in the Sgr stream are almost entirely due to the Solar reflex motion. Since the method requires fitting the Sgr stream to spatial and velocity data to predict its six-dimensional phase space configuration, it relies heavily on the shape of the Galactic potential, which also involves the value of the Galactocentric radius of the Sun ($R_{\odot}$). Moreover, the method only constrains the $V$ component of the Sun's motion. They estimate being able to recover the Solar velocity to within $10\kms$ (using  data of the quality that was expected from NASA's former Space Interferometry Mission project, \citealt{NASA_SIM2008}, which aimed to measure trigonometric parallaxes to an accuracy of 4$\mu$as).

In contrast to these previous studies, here we do not attempt to present physical models of one or more of the Galaxy's stellar streams, but rather, we develop an algorithm that is based entirely on simple geometry, and that can be applied to a sample of streams.
The new alternative method to measure the Sun's motion that we present in this paper is not correlated with the value of $R_{\odot}$, it saves us from having to analyse observations of densely populated regions in the centre of the Galaxy, it requires only 5D phase-space information about stream stars (radial velocity constraints are not needed) and does not invoke the need to model the gravitational potential of the Milky Way or the stellar populations of the disk.

The outline of this paper is as follows. In Section \ref{Method} we present the method employed in our study. Section \ref{Estimation of the Sun's Velocity using Stream models} presents our methodology to measure the Sun's velocity using (\S\ref{Employing Perfect Orbit Models}) perfect orbits and (\S\ref{Employing N-body simulated stream models}) N-body tidal stellar stream models, demonstrating the success of the method. Section \ref{Systematic bias in distance} discusses the deviation of Sun's velocity from its true value given by a systematic bias in distance measurements of stream stars. Finally in Section \ref{DISCUSSION AND CONCLUSIONS} we present the conclusion of this study.

\section{Method}\label{Method}

Our approach makes use of the assumption that for a thin stream (originating from a low mass progenitor)  all the stars lie close to a single test-particle orbit (see, e.g. \citealt{Dehnen2004thinorbit}). In general, the stars in a tidal stream have different energies, but the approximation that stream stars trace the same orbit is admissible for thin streams from low-mass progenitors.

\begin{figure}
\begin{center}
\includegraphics[angle=0, viewport= 1 3 525 400, clip, width=\hsize]{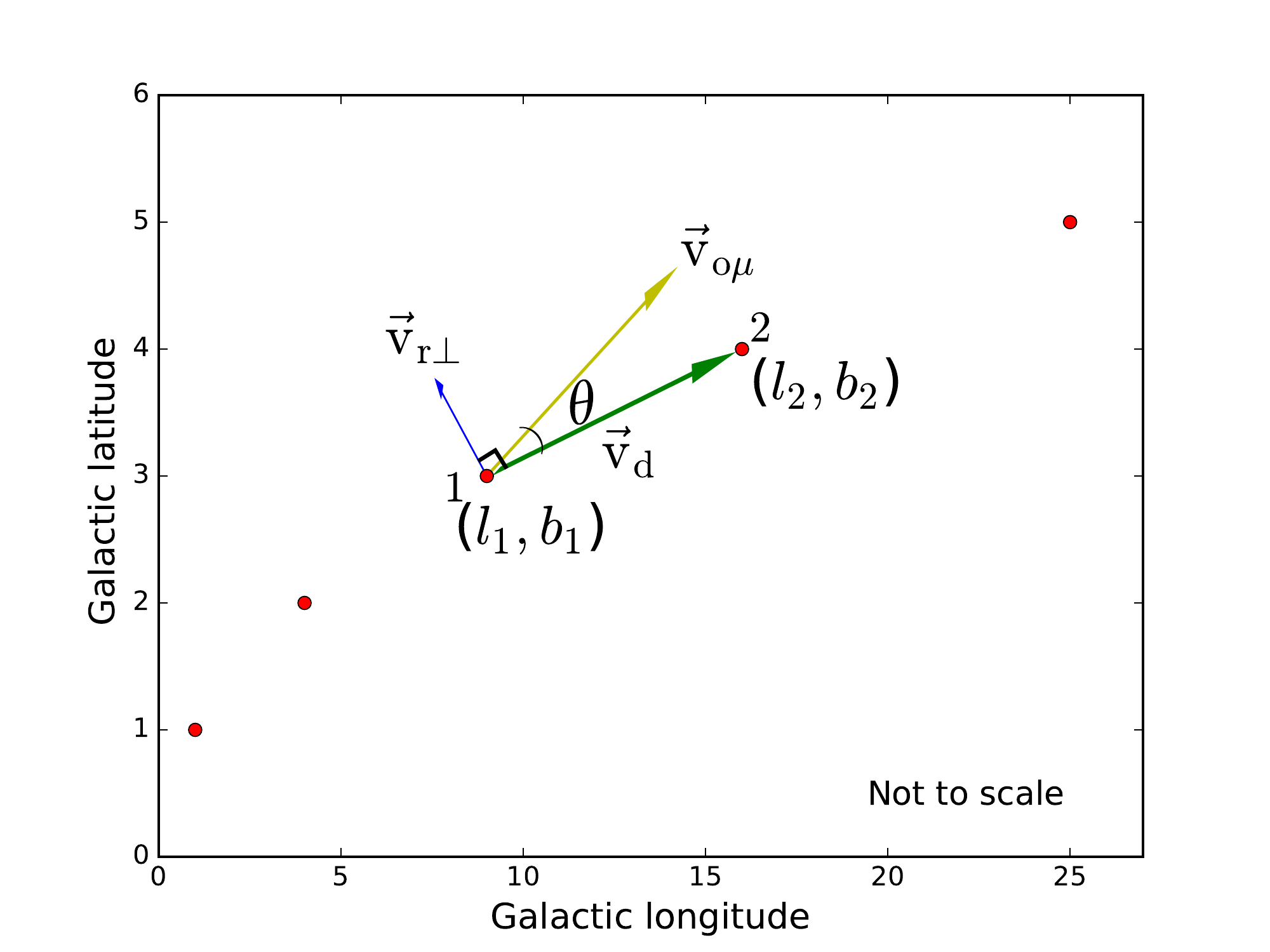}
\end{center}
\caption{Vector diagram. Red dots represent the positions at successive (equal interval) timesteps along a tiny segment of an orbit. $\bmath{v}_{d}$ is the vector that measures the path along the orbit. The proper motion vector $\bmath{v}_{o \mu}$ that gets measured in observations should lie along $\bmath{v}_{d}$ in a non-rotating frame in the Galaxy. But due to the reflex motion of the Sun, the perpendicular vector $\bmath{v}_{r \bot}$ emerges, causing the deviation of $\bmath{v}_{o \mu}$ from vector $\bmath{v}_{d}$.}
\label{fig:Conceptual figure}
\end{figure}

Consider a small segment of an orbit on the Galactic sky (as shown in Figure~\ref{fig:Conceptual figure}). The red points represent the positions of the stars (members of some stream) along their orbital structure. These points can also be viewed as different time positions for a given orbit.  We define a tangent vector $\bmath{v}_{d}$ which locally gives the direction of motion of the star's  orbit on this 2D Galactic sky. This vector is generated by connecting position at time `1' to position at time `2', along the direction of motion of the orbit. Vector $\bmath{v}_{d}$ is then given by:
\begin{equation}
	\bmath{v}_{d} = \cos(b_{1}) (\ell_{2}-\ell_{1}) \bmath{\hat{\ell}} + (b_{2}-b_{1}) \bmath{\hat{b}} \, ,
    \label{eq:measure structure}
\end{equation}
where
$\bmath{\hat{\ell}}$ is the unit Galactic longitude vector and $\bmath{\hat{b}}$ is the unit Galactic latitude vector.

Our assumption that stellar streams follow the orbit of the constituent stars means that the path depicted in  Figure~\ref{fig:Conceptual figure} can be recovered from the position of the stream on the sky, so that the two time intervals `1' and `2' along the path can be equivalently thought of as two stars `1' and `2' along the orbit.

With this assumption, in the Galaxy's non-rotating frame, the observed proper motion vector of star `1' (see  Figure~\ref{fig:Conceptual figure}) should align along the vector $\bmath{v}_{d}$, since `1' must practically trace out the orbit of the succeeding star `2' (by the definition of an orbit). But due to the motion of the observer's frame in the Galaxy, the observed proper motion vector is different in direction and magnitude. We define this observed proper motion vector $\bmath{v}_{o \mu}$ as:
\begin{equation}
\bmath{v}_{o \mu}= \mu_{l_{1}} \cos(b_{1}) \bmath{\hat{\ell}} + \mu_{b_{1}} \bmath{\hat{b}} \, .
\end{equation}
Therefore, the perpendicular component (which we call $\bmath{v}_{r \bot}$) of vector $\bmath{v}_{o \mu}$ to $\bmath{v}_{d}$ emerges totally due to the reflex motion of the Sun as seen at position `1' and is given by:
\begin{equation}
\bmath{v}_{r \bot} = (|\bmath{v}_{o \mu}| \, \sin\theta ) \bmath{\hat{v_{r}}_{\bot}} \, ,
\end{equation}
where $\theta$ is the angle between $\bmath{v}_{d}$ and $\bmath{v}_{o \mu}$ and $\bmath{\hat{v}}_{r \bot}$ is the unit vector normal to $\bmath{v}_{d}$.

However, even for a simple orbit, the precession of the orbital plane in the Galaxy will also contribute to the vector $\bmath{v}_{r \bot}$. We can estimate approximately the contribution of precession to $\bmath{v}_{r \bot}$, using the analytic approximation of \citet{Steiman1990precession}. Their formula is valid for a very simple case assuming a circular orbit evolving in a spheroidal potential in which the reference frame is not tumbling. The precession rate is then:
\begin{equation}
\dot{\Omega}_{p}=-\, \frac{3 \Phi_{2}(r)}{2 \, r \, v_{circ \odot}}\cos\,i \, ,
\end{equation}
where
\begin{equation}
\Phi_{2}(r) = \frac{v_{circ \odot}^{2}}{2} \left(\frac{1-q_{\phi}^{2}}{q_{\phi}^{2} + \frac{1}{2}}\right) \, .
\end{equation}
Here $\Omega_{p}$ is the longitude of the ascending node of an orbit, $i$ is the inclination of the orbital plane, $\Phi_{2}(r)$ is one of the components of the expansion of the scale-free logarithmic potential function $\Phi(r)$, and $q_{\phi}$ is the (spheroidal) flattening of the potential. Taking an orbit at a typical radius of $r=30\kpc$ in a potential with density flattening of $q_{\rho} = 0.8$ and with circular velocity of $v_{circ \odot} = 200\kms$, yields $\mid\dot{\Omega}_{p}\mid\,=\, 0.11 \cos\,i\,\masyr$. The component along the vector $\bmath{v}_{r \bot}$ then becomes $\bmath{v}_{r \bot}^{'} = 0.055 \cos(2i-90) \masyr$, i.e. with a maximum at $i=45\deg$ of $\mid\dot{\Omega}_{p}\mid\,=\,0.055\masyr$. This corresponds to a maximum of 4\% of the proper motion of the corresponding circular orbit. This estimate shows that the effect of precession should be relatively small. Note also that if we consider multiple streams that are on different orbits, then they will also have different orbital plane inclinations $i$. Hence the precession corrections will tend to cancel out on average.

The procedure we follow is to sample different values of the Sun's three-dimensional velocity $\bmath{\textbf{V}}_{\odot}$ = ($u_{\odot}$, $v_{\odot}$, $w_{\odot}$) using a Markov Chain Monte Carlo algorithm. The apparent stream motion is calculated as the reflex motion vector $\bmath{v}_{r \bot}$ and is compared against $\bmath{v}_{r \bot}$ obtained from data. The figure of merit we adopt is the likelihood of the data given the stream model. In Section \ref{Employing Perfect Orbit Models} below we investigate first the results given a set of perfect orbit streams, while in Section \ref{Employing N-body simulated stream models}, we will take the more realistic case of  an N-body stream due to tidally disrupted satellites.

For these calculations we will make use of the realistic Galactic potential model of \citet{Dehnen1998Massmodel} (their model {\bf{1}}), which contains a bulge, thin disc, thick disk, interstellar medium, and a halo component. We stress that this potential model is only used to set up the artificial stream realisations, and is in no way used to deduce the Solar velocity vector. The method we present here is independent of any models of the Galactic potential.

\section{Estimation of the Sun's Velocity using Stream models}\label{Estimation of the Sun's Velocity using Stream models}

\subsection{Employing Perfect Orbit Models}\label{Employing Perfect Orbit Models}

In order to be completely assured of our method, we first demonstrate a proof of concept, using 
perfect orbit models (which in principle can be considered as an ideal stream case). Since orbits are infinitely thin curves, the Sun's velocity $\bmath{V}_{\odot}$ should be perfectly recovered to within the biasses created by the orbital precession.

\begin{figure}
\begin{center}
\vbox{
 \begin{overpic}[scale=0.40]{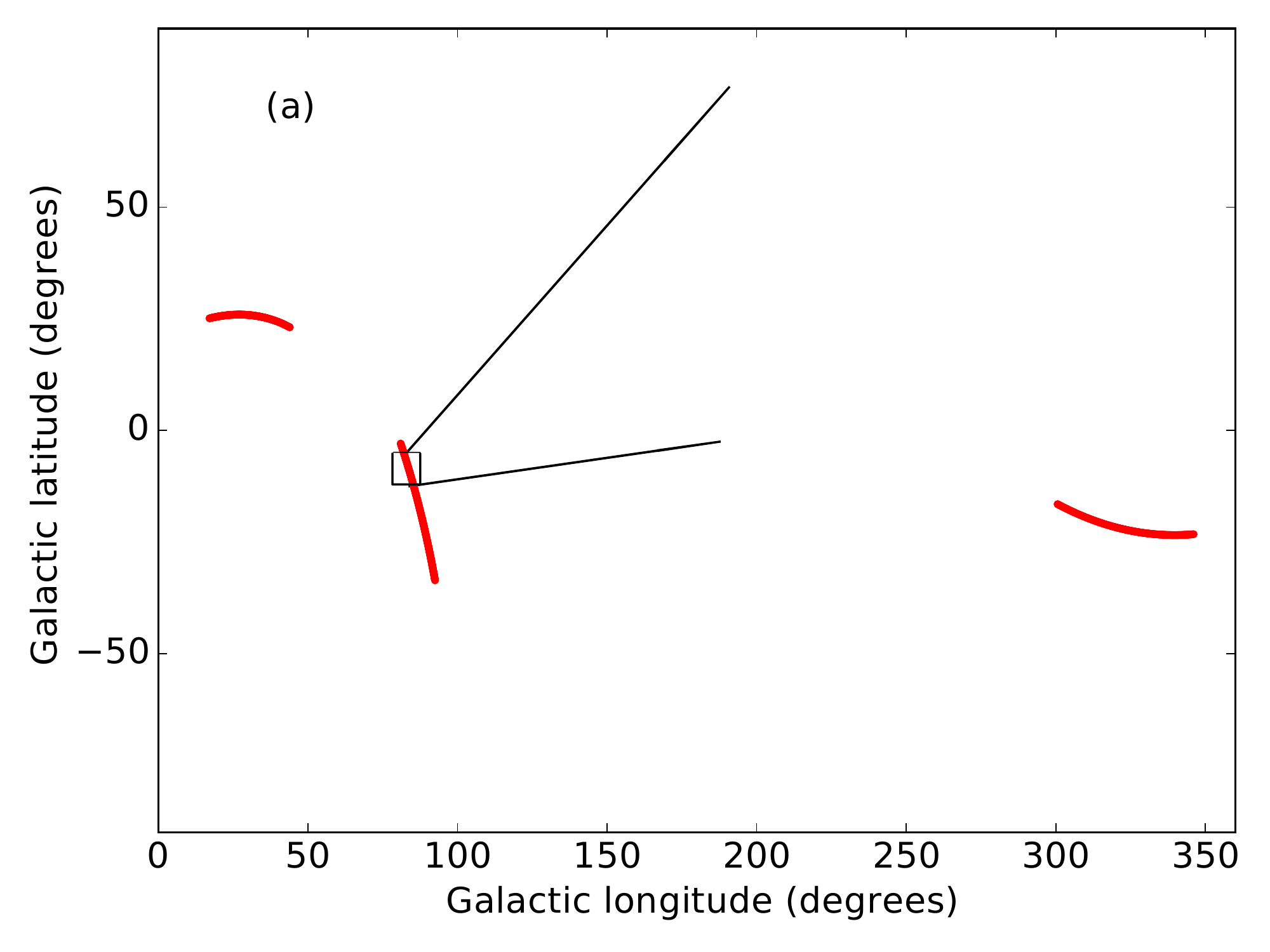}
  \put(53,38){\includegraphics[angle=0, viewport= 0 0 1580 1680, clip, width=\hsize]{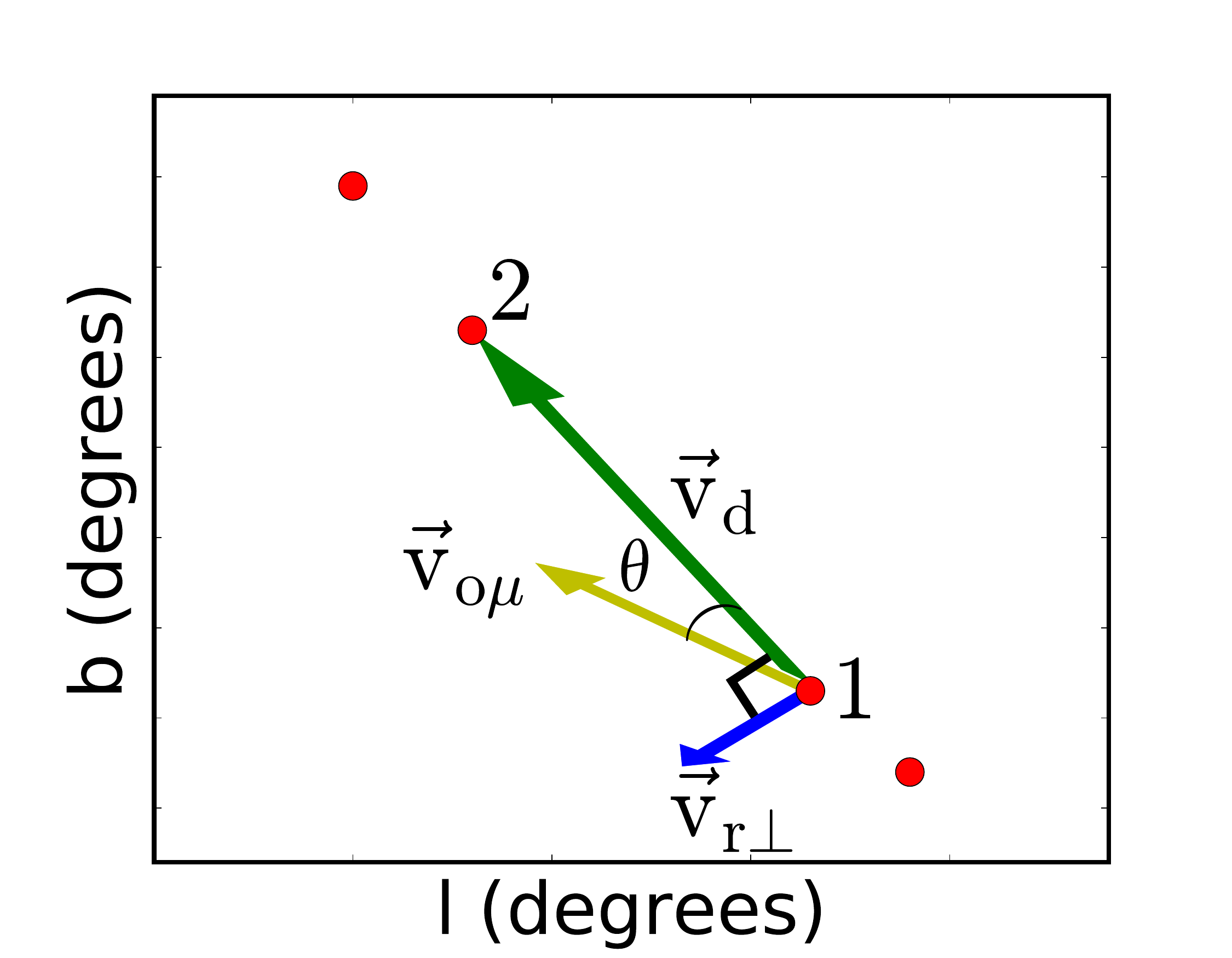}}
 \end{overpic}
\includegraphics[angle=0, viewport=1 3 570 480, clip, height=6.5cm]{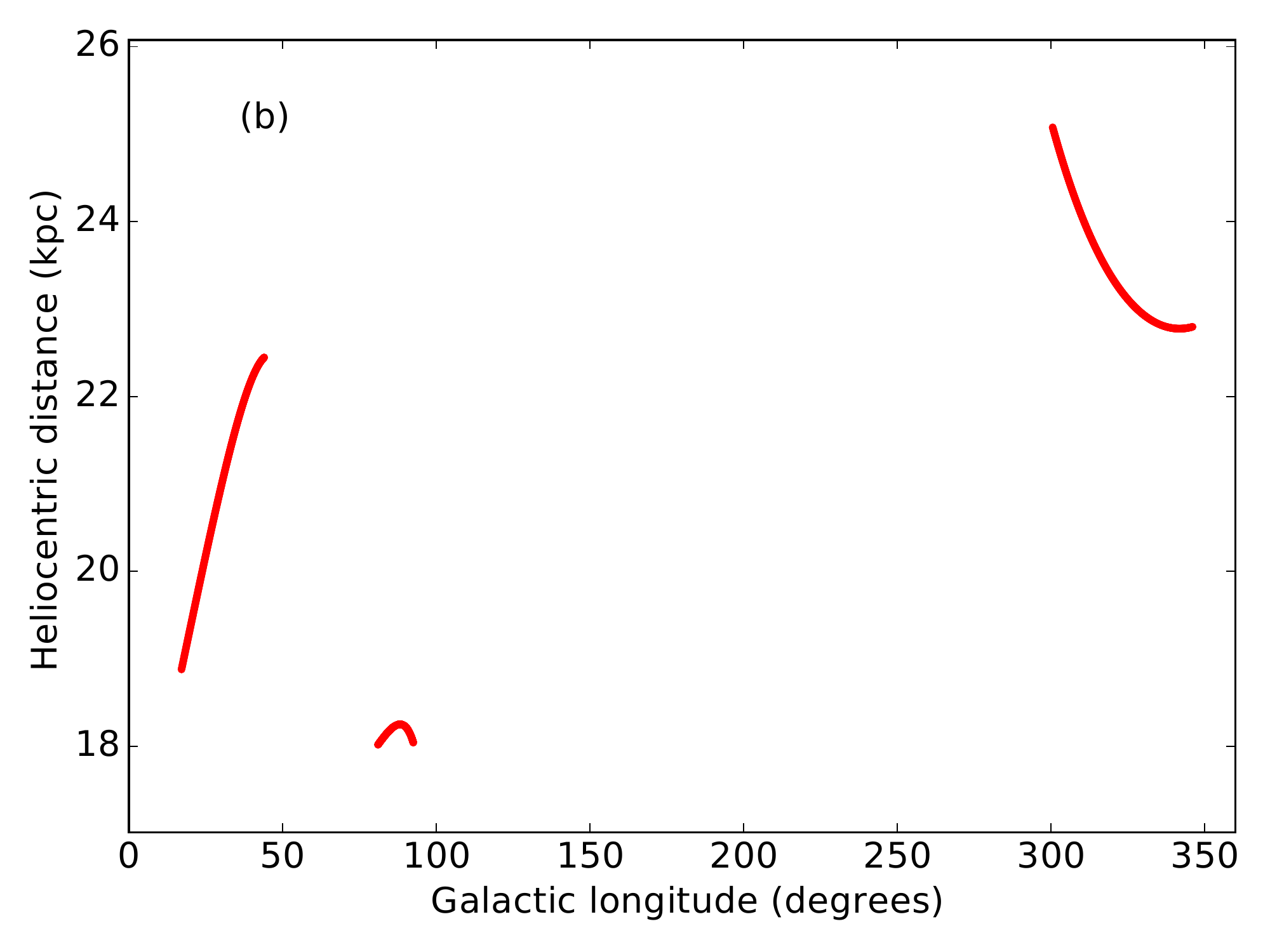}
}
\end{center}
\caption{Sky view of the perfect orbits. (a) shows the path of the orbits on the Galactic sky and (b) represents the Heliocentric distances that these orbits span. The orbits were integrated in the Galactic potential model \textbf{1} of \citet{Dehnen1998Massmodel}. The zoomed-in panel in (a) represents a small segment of the orbit detailing the geometry of our procedure. It is the vector $\bmath{v_{r \bot}}$ that our model is compared against.}
\label{fig:Perfect Orbit Sky view}
\end{figure}

To achieve this, we selected three 6D phase space positions drawn randomly to give the orbits' initial conditions. Each of these initial conditions were then integrated for $T = 0.06\Gyr$ in the Galactic potential model described above to form an orbit (the value of $T$ was chosen, somewhat arbitrarily, just so that the orbits appear long enough to mimic observed streams found in the SDSS). Since we need to constrain 3 components of the Sun's velocity, we either need a single stream that probes  different regions of the sky in such a way that each component of the reflex motion dominates, or we need a minimum collection of 3 stream segments which again explore the sky so that the corresponding reflex motion components are significant. The latter possibility is considered here, since all low-mass globular cluster streams currently known are approximately great-circle segments, at most a few tens of degrees long. Once integrated, the complete phase-space information of these 3 orbits was then transformed from the Galactocentric Cartesian frame to a Heliocentric (observable) frame using the Sun's parameters (that we refer to as the \textit{true} parameters) as ($R_{\odot}$, $\textbf{V}_{\odot}$) = ($R_{\odot,{\rm T}}$, $u_{\odot,{\rm T}}$, $v_{\odot,{\rm T}}$, $w_{\odot,{\rm T}}$) = ($8.34\kpc$, $9.0\kms$, $255.20\kms$, $7.0\kms$). However, only 5D information was retained in the form of ($\ell$, $b$, $d_{\odot}$, $\mu_{\ell}$, $\mu_{b}$). The resulting spatial projection of the randomly-chosen orbits is shown in  Figure~\ref{fig:Perfect Orbit Sky view}.

A Markov Chain Monte Carlo algorithm is used to survey the parameter space of Solar velocity components ($u_{\odot}$, $v_{\odot}$, $w_{\odot}$), where the model likelihood is taken to be:
\begin{equation}
\begin{split}
L[u_{\odot},v_{\odot},w_{\odot}] = 
\sum_{\rm Data} -\ln(\sigma_{\ell} \, \sigma_{b}) \\
-\Big(\dfrac{ {\mu_{\bot, \ell}^{\rm data}} - {\mu_{\bot, \ell}^{\rm model}}}{\sqrt[]{2} \sigma_{\ell}} \Big)^2
-\Big(\dfrac{ {\mu_{\bot, b}^{\rm data}} - {\mu_{\bot, b}^{\rm model}}}{\sqrt[]{2} \sigma_{b}} \Big)^2 \, ,
\end{split}
\end{equation}
where $\mu_{\bot, \ell}^{\rm data}$ and $\mu_{\bot, b}^{\rm data}$ are the observed $\ell,b$ components of $\bmath{v}_{r \perp}$, and $\mu_{\bot, \ell}^{\rm model}$ and $\mu_{\bot, b}^{\rm model}$ are the corresponding model predictions. In Section~\ref{Employing N-body simulated stream models} below,
$\sigma_{l}$ and $\sigma_{b}$ will represent proper motion uncertainties of the stream stars in, respectively, the Galactic longitude and latitude directions. However, for the perfect orbit model tests, we allow the MCMC algorithm to fit a global value for these two dispersion parameters (in this situation, they can be considered as model mismatch errors).

\begin{figure}
\begin{center}
\includegraphics[angle=0, viewport= 1 3 525 520, clip, width=\hsize]{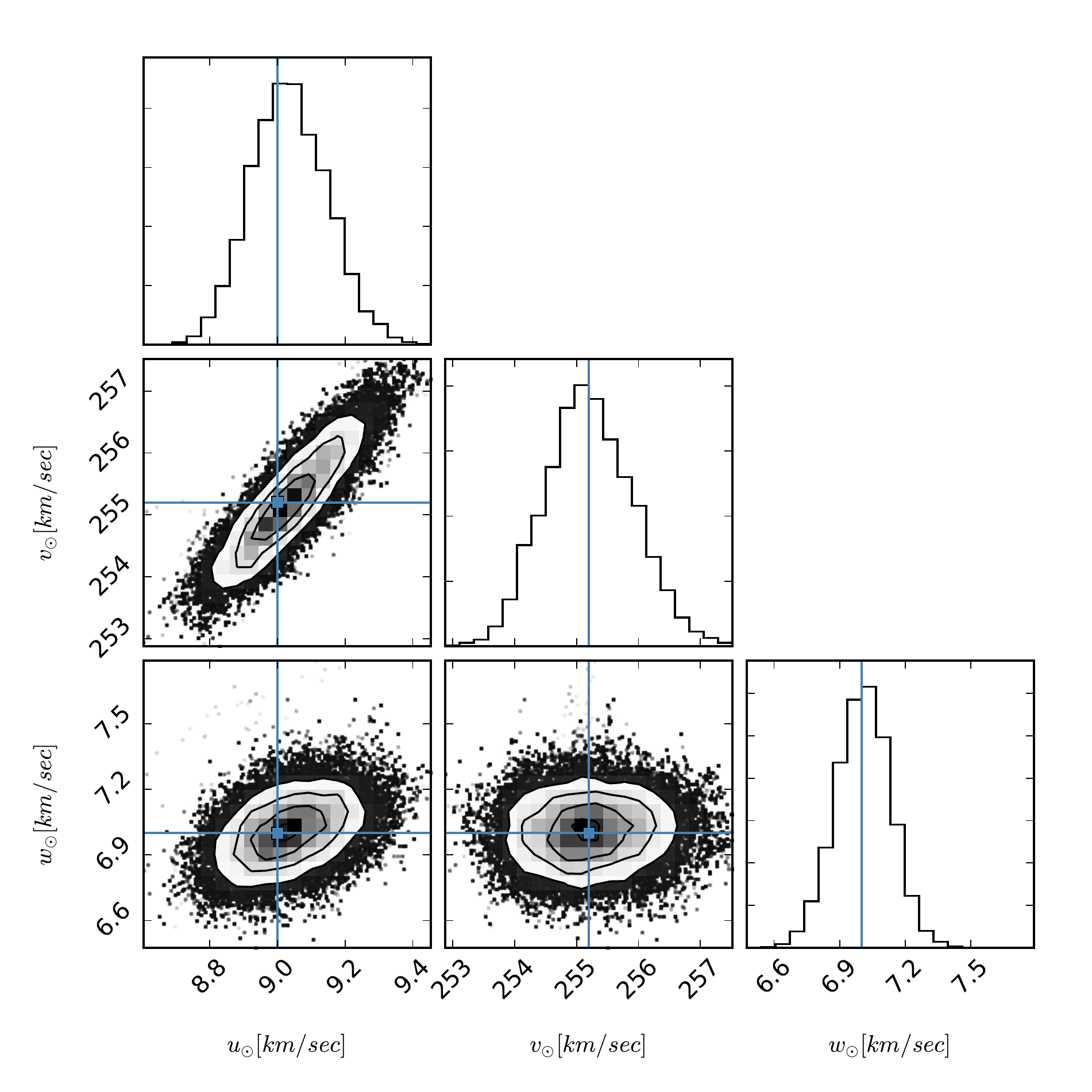}
\end{center}
\caption{Correlation function plot for the perfect orbit test. The panels represent the probability distribution function and parameter-parameter correlations of the Sun's velocity components obtained through the application of the MCMC algorithm. The blue lines represent the true input values of the Sun's velocity. }
\label{fig:Corner Plot}
\end{figure}

Figure~\ref{fig:Corner Plot} shows the resulting distribution of Solar velocity components explored by the MCMC algorithm in $1.5 \times 10^6$ iterations in the form of a triangular correlation diagram. The most likely values are found to be ($u_{\odot}$, $v_{\odot}$, $w_{\odot}$) = (9.03, 255.26, 7.001) $\kms$ which shifts the measured values from the true values by ($u_{\odot}-u_{\odot,{\rm T}}$, $v_{\odot}-v_{\odot,{\rm T}}$, $w_{\odot}-w_{\odot,{\rm T}}$) = (0.03, 0.06, 0.001) $\kms$, and the corresponding uncertainties ($\sigma_{u}$, $\sigma_{v}$, $\sigma_{w}$) = (0.11, 0.68 , 0.13) $\kms$. Thus the results from this idealised example clearly establish the proof of concept. We next test if the method works on more physical stellar stream systems and if these could actually be used to constrain the Solar motion in the Galaxy.

\subsection{Employing N-body simulated stream models}\label{Employing N-body simulated stream models}

In reality, star streams form from the tidal disruption and dissolution of satellites. The escaping stars need to be lifted out of the potential well of their progenitor, and in so doing, they end up with different energies (and hence on different orbits) than their progenitor. Thus to obtain a more realistic description of streams, we decided to produce a set of N-body models in the \citet{Dehnen1998Massmodel} Galactic potential model \textbf{1}. For this, we used the GyrafalcON N-body integrator \citep{Dehnen2000GyrafalcON} from the NEMO software package \citep{Teuben1995NEMO}.

The initial phase space distribution of the progenitors of the streams was selected as follows. The initial position of each satellite was drawn at a random direction as seen from the Galactic Center, and with a uniform probability of lying in the Galactocentric distance range of $[10\textup{--}30]\kpc$. The mean velocity of each satellite was selected randomly from an isotropic Gaussian distribution with (one-dimensional) dispersion of $100\kms$ \citep{Harris1976,Bosch1999}. At these phase space positions, each progenitor was constructed using a King model (\citealt{King1966}). The mass, tidal radius and ratio between central potential and velocity dispersion were sampled uniformly between the ranges $M_{sat} = [2\textup{--}5] \times 10^{4}\msun$ , $r_{t} = [20\textup{--}80]\pc$ and $W_{sat} = [2\textup{--}4]$. 

Somewhat arbitrarily, we chose to model a set of 22 streams. At present, $\sim 9$ low-mass streams of probable globular cluster progentors are known within the $\sim 1/4$ of the sky in the North Galactic SDSS footprint: Acheron, Lethe, Cocytos, Styx, Hermus, Hyllus, Palomar~5, NGC 5466, and GD-1 (see, e.g.  \citealt{Grillmair:2016ju}). An additional 6 narrow streams (Ophiuchus, PS1-A,PS1-B, PS1-C, PS1-D and PS1-E) were discovered in the $\sim 3/4$ of the sky covered by the Pan-STARRS survey \citep{Bernard2014panstarstream, 2016MNRAS.463.1759B}. We expect several more to come to light thanks to the Gaia survey, which will cover the full sky and allow for de-contamination of foreground populations by proper motion. Hence a choice of $\sim 20$ systems for our sample of streams is a conservative estimate of what should be well-measured by Gaia within a few years.

\begin{figure*}
\begin{center}
\vbox{
\hbox{
\hskip 1.0cm
\includegraphics[angle=0, viewport= 0 0 625 530, clip, height=6.5cm]{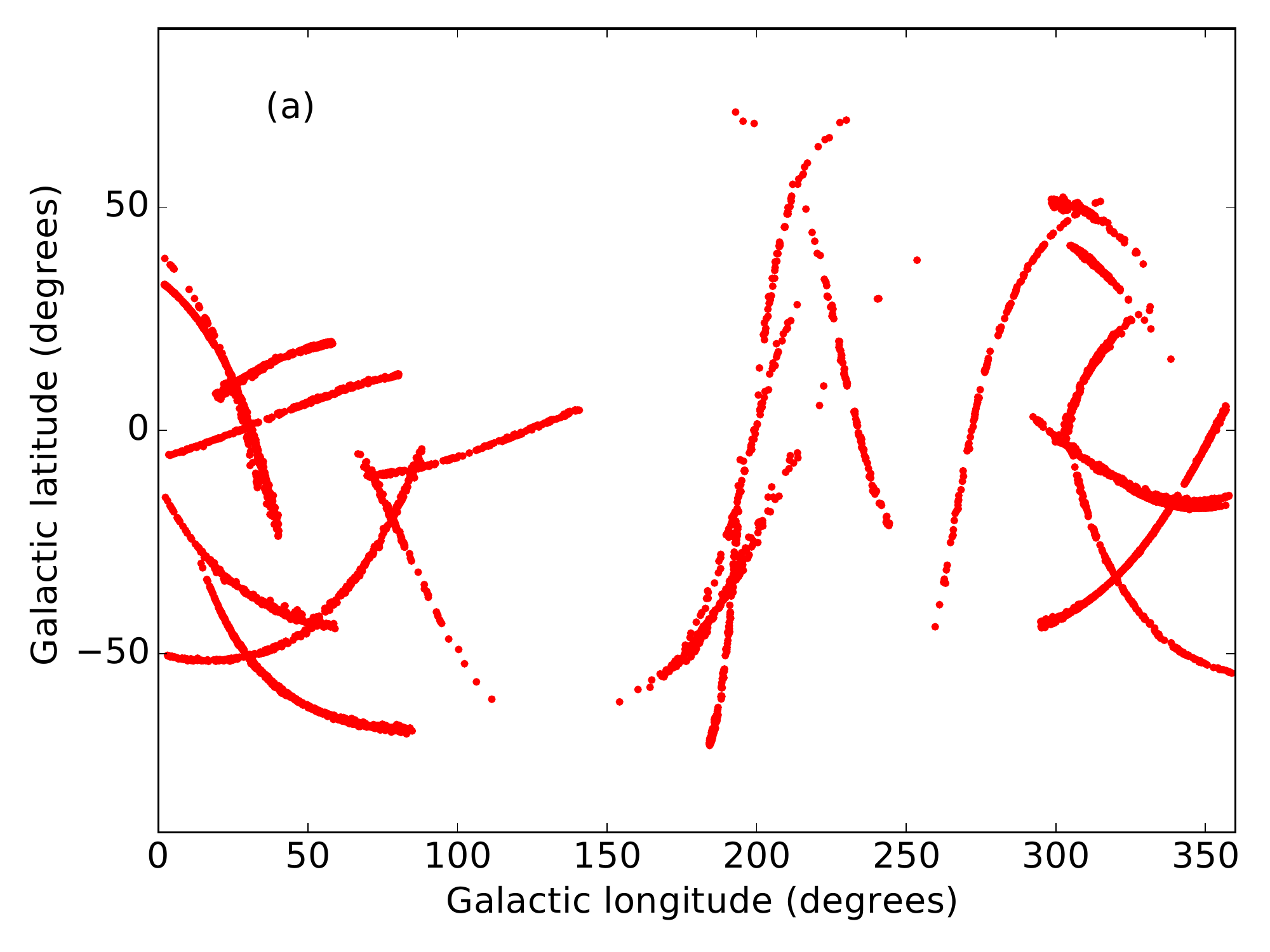}
\includegraphics[angle=0, viewport= 0 0 625 530, clip, height=6.5cm]{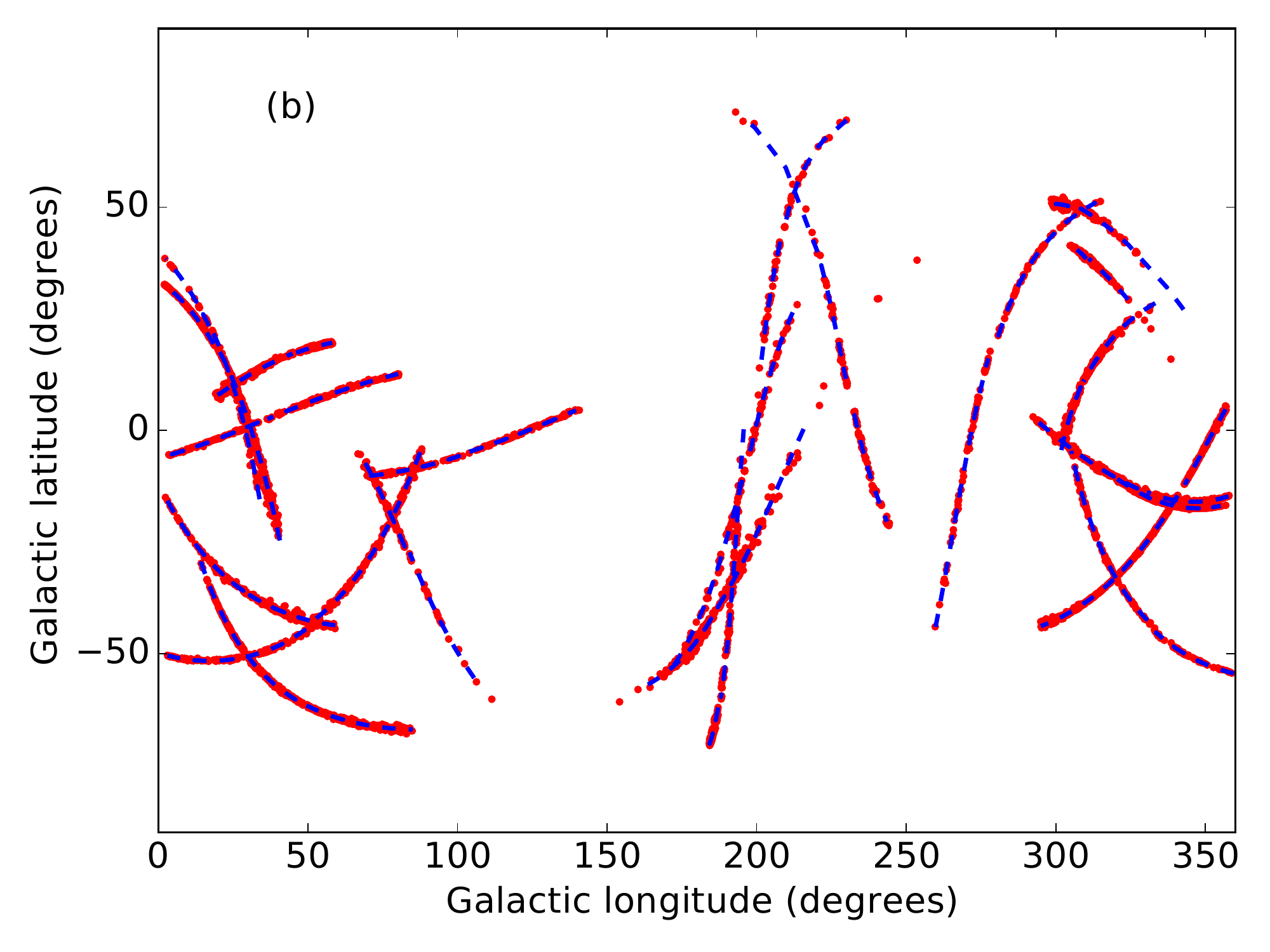}
}
\hbox{
\hskip 1.0cm
\includegraphics[angle=0, viewport= 0 0 625 530, clip, height=6.5cm]{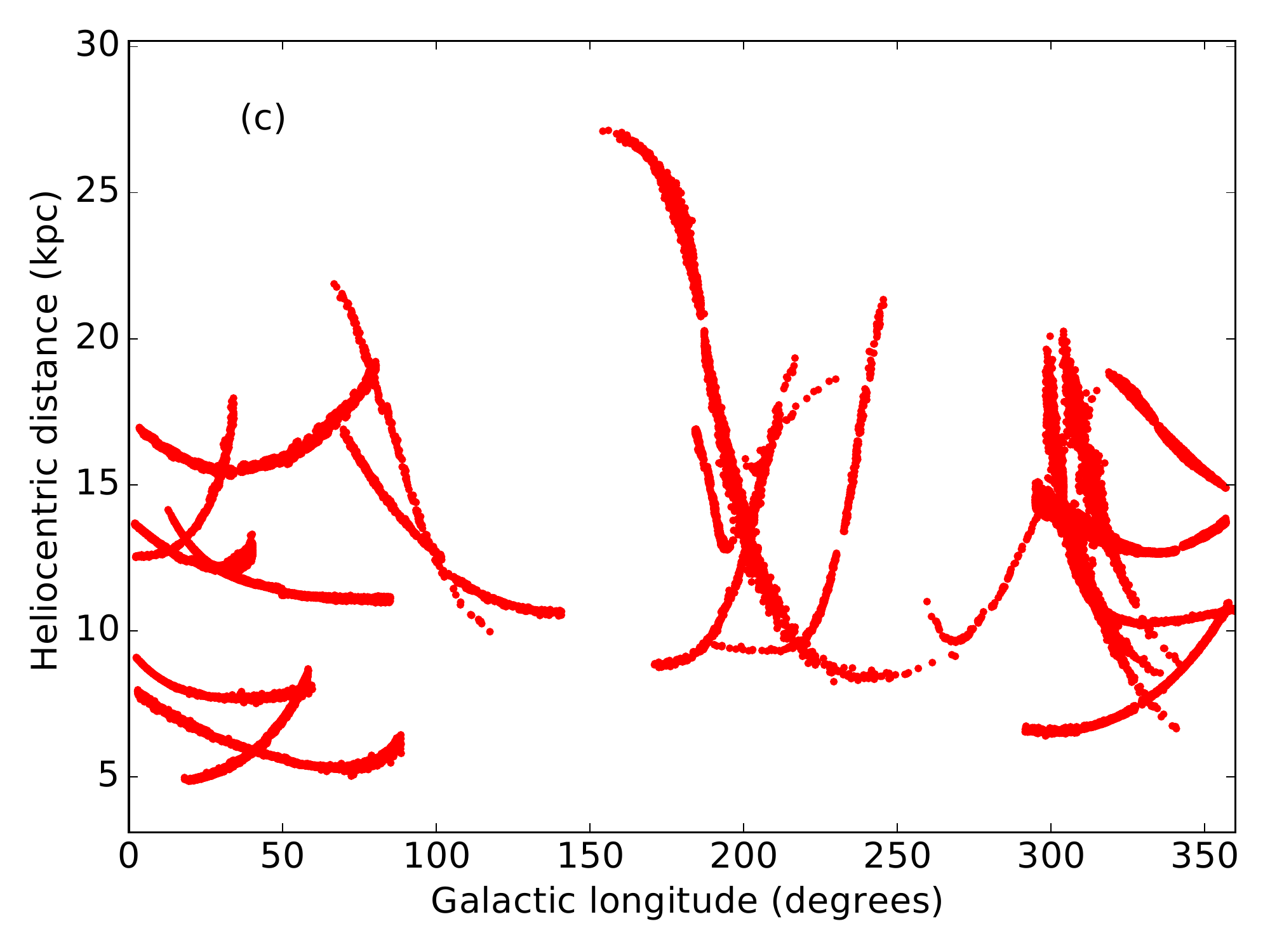}
\includegraphics[angle=0, viewport= 0 0 625 530, clip, height=6.5cm]{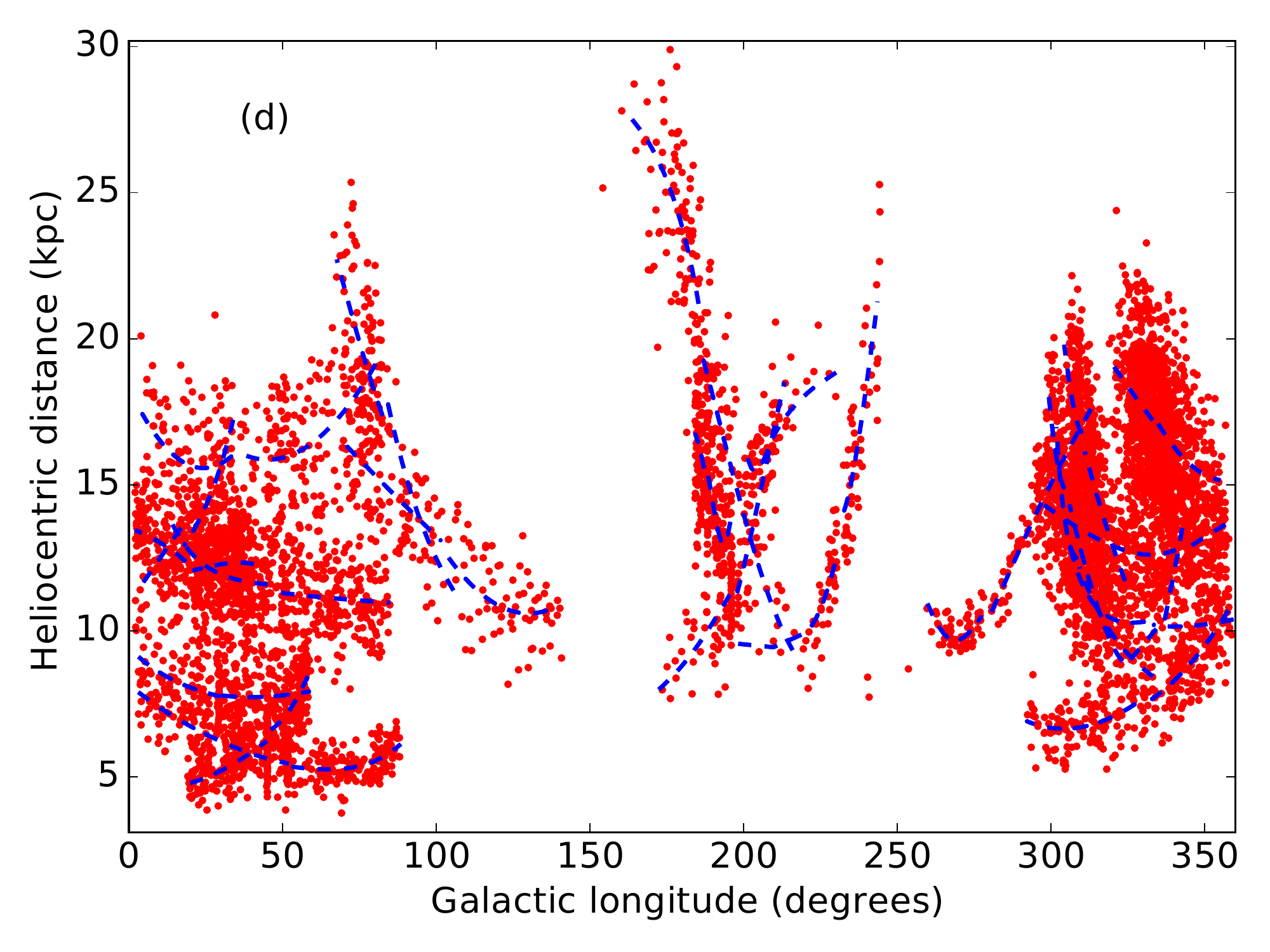}
}
}
\end{center}
\caption{Sky view of the N-body tidal stellar streams. The left panels show the positions in Galactic sky (a) and Heliocentric distance (c) that the simulated streams span. The right panels display the same information once the errors in proper motion and distance are applied (consistent with Gaia and CFIS  survey uncertainties). The red points are the mock stellar points and the blue curves are the fitted curves to these data points. On average, we keep 230 stellar data points per stream.}
\label{fig:Sky view of tidal simulated streams}
\end{figure*}

Once the progenitors were initialised in phase space, they were then evolved independently over a time period between $[2\textup{--}8]\Gyr$ in the same Galactic mass model mentioned above. During this period of time, most of the progenitors were tidally disrupted, giving rise to streams. Those that  did not disrupt were re-sampled and evolved. The simulated streams were then transformed from the Galactocentric to the Heliocentric frame (using the \textit{true} parameter set for the Sun) and again only the 5-D phase-space information was preserved. Figure~\ref{fig:Sky view of tidal simulated streams}a represents the Galactic sky structure of these tidal stream models. While we of course do not as yet know the stream discoveries that will be made with the Gaia DR2 catalogue, the distribution shown in Figure~\ref{fig:Sky view of tidal simulated streams} does not appear to be implausible.

To make a fair comparison, the stream models were degraded with realistic uncertainties.
\begin{itemize}
\item \textit{Uncertainty in observed proper motions}: We introduced End of mission Gaia  uncertainties for the proper motions into the simulated data. The Gaia errors depend upon the colour and magnitude of the stars. For this, we needed to assign magnitudes to the mock data points. This was implemented using the Padova stellar population models \citep{Marigo2008Padova}. A star in the mock stream was selected and its absolute magnitude (${\rm M_g}$) was drawn in the g-band from the isochrone metallicity ${\rm [Fe/H]=-1.5}$ and age $10\Gyr$, appropriate for a halo globular cluster. Using perfect distance information of this stellar point and the absolute magnitude, an apparent magnitude was assigned to every star. Using the colour transformations detailed in \citet{GaiaJordi2010}, we converted the magnitude to the Gaia G-band, and limited these to $G=20.5$. Once the magnitude value was assigned, the uncertainty in proper motion ($\mu_{\ell}$, $\mu_{b}$) was generated using the ``End-of-Mission Sky Average Astrometric Performance Chart''\footnote{Available on Gaia's official website {\tt https://www.cosmos.esa.int/web/gaia/sp-table1}}. We also assumed a minimum stream velocity dispersion of $5\kms$, which is converted into proper motion and added in quadrature to the observational uncertainties.
\item \textit{Uncertainty in distance measurements}: We also introduce a 10\% uncertainty error in the heliocentric distance ($d_{\odot}$) measurements to the model stream stars. The motivation for this is that although Gaia parallaxes will be excellent for bright nearby stars, the majority of Galactic halo tracers will lie near Gaia's faint detection limit, with no geometric parallax information. However, photometric parallaxes will be measurable for such stars, using for instance the metallicity-magnitude-distance  calibration of \citet{2008ApJ...684..287I} which is applicable to main-sequence stars. \citet{2008ApJ...684..287I} show that 5\% distance uncertainties are achievable with this method with good photometry; our 10\% uncertainty value is chosen to be a plausible average value. The necessary photometry (in particular the u-band) is currently being obtained in the Northern hemisphere as part of the Canada-France Imaging Survey\footnote{\tt http://www.cfht.hawaii.edu/Science/CFIS/} and starting in $\sim 2021$ will also be available in the Southern hemisphere thanks to the LSST. 
\end{itemize}

After this procedure, the simulated stream particles get smeared out in phase space, as shown in Figure~\ref{fig:Sky view of tidal simulated streams}. However, we need the streams to be approximated by a curve along which the vector $\bmath{v}_{d}$ can be calculated over the full length of the stream. It is thus necessary to curve-fit the stream data. We implemented this using a simple quadratic polynomial function. The fitting procedure was conducted only in the (two-dimensional) sky frame in a coordinate system similar to the Galactic system, but rotated to ensure that both arms of the stream run, as closely as possible, to the equator of the new rotated coordinate frame (this is only approximate, since in general, streams do not follow precisely great circle paths). We define the coordinates of the new rotated frame to be $\ell_{new}$ and $b_{new}$. We fitted for $b_{new}$ and $d_{\odot}$ in the transformed coordinate system ($l_{new}$, $b_{new}$) using a Singular Value Decomposition (SVD) algorithm with polynomial functional form:
\begin{equation}
b_{new} = a_{1} + b_{1} \ell_{new} + c_{1} \ell_{new}^2
\end{equation}
and
\begin{equation}
d_{\odot} = a_{2} + b_{2} \ell_{new} + c_{2} \ell_{new}^2 \, ,
\end{equation}
where $a_{i}$, $b_{i}$ and $c_{i}$ are the fitting parameters. The 2 arms in all of these 22 streams were fitted independently. Once the fitting procedure was complete, the fitted curves were then transformed back to Galactic coordinates. Figure~\ref{fig:Sky view of tidal simulated streams} (right panels) represents the streams with uncertainties introduced along with the  curves fitted to them. Once fitting is done, equations (7) and (8) are then used to calculate the vector $\bmath{v}_{d}$ and the heliocentric distances at every stellar point. $\theta$ is still the angle between $\bmath{v}_{d}$ and $\bmath{v}_{o \mu}$, where $\bmath{v}_{o \mu}$ is the observed proper motion vector.

\begin{figure}
\begin{center}
\includegraphics[angle=0, viewport= 1 3 525 520, clip, width=\hsize]{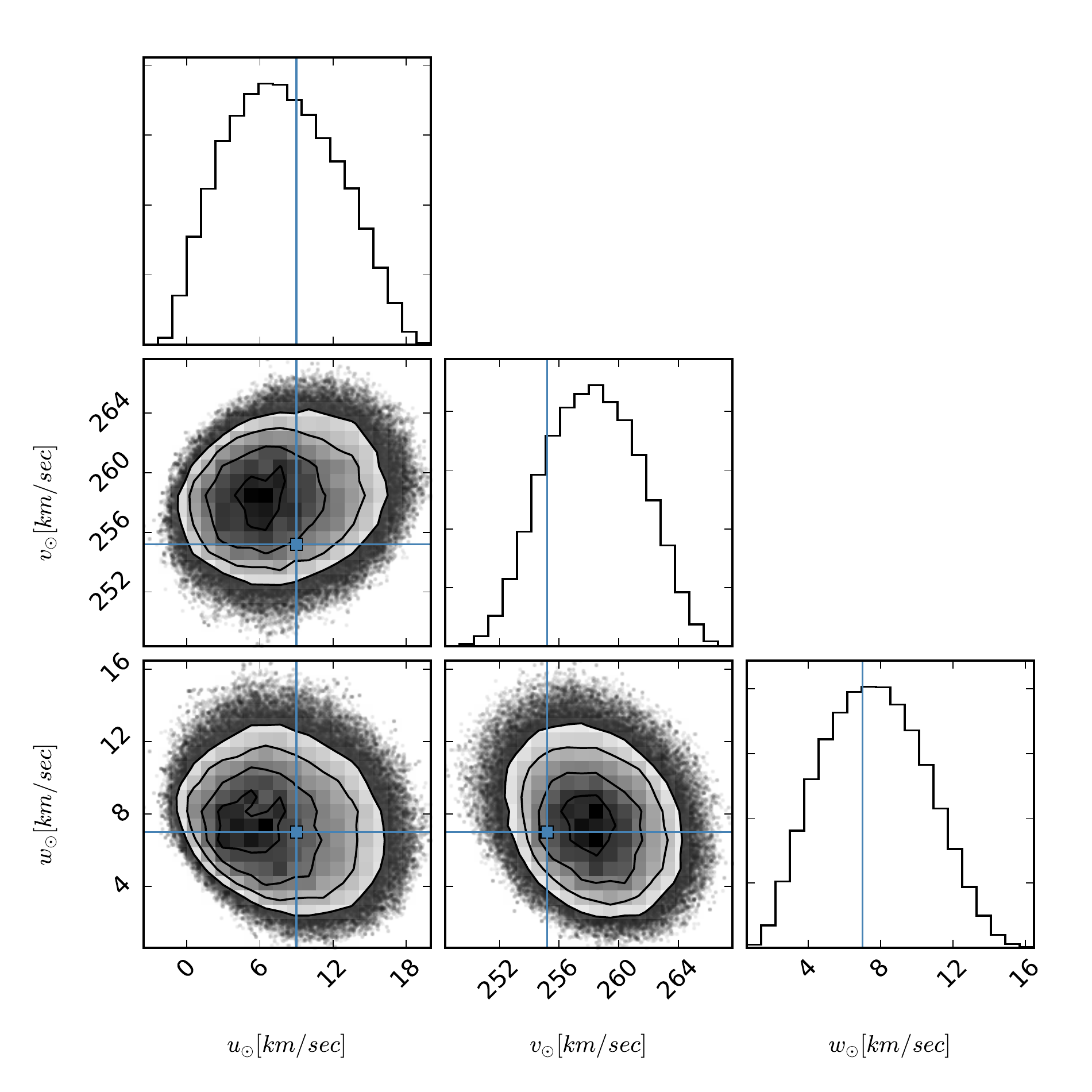}
\end{center}
\caption{Correlation function plot for the N-body stream tests. The model distributions of the $u_{\odot}$, $v_{\odot}$ and $w_{\odot}$ components of the Sun's velocity, are shown, as predicted by our method in $2.0 \times 10^5$ MCMC iterations. The input values that were used for the Sun's velocity are displayed with the blue lines and are ($u_{\odot}$, $v_{\odot}$ and $w_{\odot}$ )= (9.0, 255.2, 7.0 ) $\kms$. The MCMC method is clearly able to recover these values to useful accuracy from the stream kinematics.}
\label{fig:Corner plot for streams}
\end{figure}

In this case, Sun's Galactic velocity solution (shown in Figure~\ref{fig:Corner plot for streams}) was recovered as : ($u_{\odot}$, $v_{\odot}$, $w_{\odot}$) = (7.80, 258.25, 7.69) $\kms$. This means that the bias estimated between the observed and true value of the Sun's velocity is ($u_{\odot}-u_{\odot,{\rm T}}$, $v_{\odot}-v_{\odot,{\rm T}}$, $w_{\odot}-w_{\odot,{\rm T}}$) = (-1.20, 3.05, 0.69) $\kms$, with uncertainties ($\sigma_{u}$, $\sigma_{v}$, $\sigma_{w}$) = (4.16, 3.04, 2.74)  $\kms$.

\subsection{Systematic bias in distance}\label{Systematic bias in distance}

\begin{figure}
\begin{center}
\includegraphics[angle=0, viewport= 1 3 525 520, clip, width=\hsize]{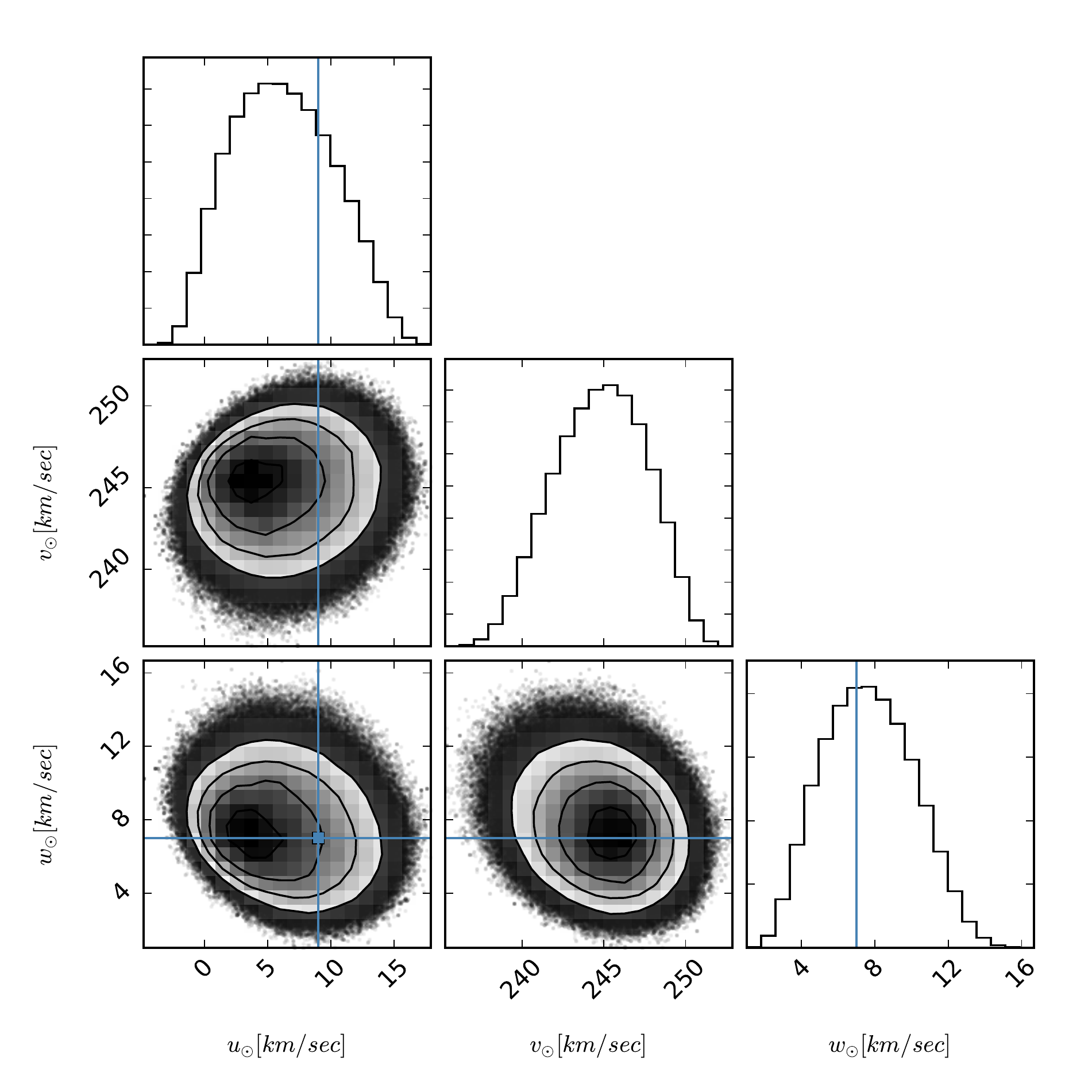}
\end{center}
\caption{As Figure~\ref{fig:Corner plot for streams}, but for an underestimated 5\% systematic bias introduced into the simulated Heliocentric distance information. In this case, we find a significant bias of $\sim 10\kms$ in the $v$ component of the Sun's motion. The MCMC algorithm was made to run for $2.0 \times 10^5$ iterations in this case.}
\label{fig:Corealtion function study for systematic bias}
\end{figure}

For completeness, we consider next what the effect of a $\pm 5$\% distance bias would have on the derivation of the Solar velocity; such a bias could arise in principle from an incorrect calibration in the photometric distances. To this end, we reran the algorithm on the simulated data and simply forced all of the stellar particles to be 5\% less distant. The resulting most likely solution has:
($u_{\odot}-u_{\odot,{\rm T}}$, $v_{\odot}-v_{\odot,{\rm T}}$, $w_{\odot}-w_{\odot,{\rm T}}$) = (-2.75, -9.69, 0.66) $\kms$, and uncertainty ($\sigma_{u}$, $\sigma_{v}$, $\sigma_{w}$) = (3.73, 2.79, 2.45) $\kms$. The correlation function plot for this case is shown in Figure~\ref{fig:Corealtion function study for systematic bias}. Re-running this test using distances that are systematically 5\% overestimated gives qualitatively similar results, but with a velocity bias of ($u_{\odot}-u_{\odot,{\rm T}}$, $v_{\odot}-v_{\odot,{\rm T}}$, $w_{\odot}-w_{\odot,{\rm T}}$) = (-0.99, 13.58, 1.56) $\kms$.

The bias in the distance measurements clearly affects the vector $\bmath{v}_{r \bot}$ (calculated at the stellar points) and hence affects the Solar velocity $\bmath{\textbf{V}}_{\odot}$ estimation. In our study (and as can be seen in Figure~\ref{fig:Corealtion function study for systematic bias}, $v_{\odot}$ panel), a large shift was observed in the $v_{\odot}$ component of the velocity of the Sun from the true $v_{\odot}$ value, but the other components were less affected. This is not specific to the technique but rather to the overall phase space distribution of stellar streams with respect to the Sun's motion. The phase space distribution and orientation of the streams in the example happened to be such that the $v_{\odot}$ component faced a higher offset from the $\textit{true}$ value than the other two components.

\section{DISCUSSION AND CONCLUSIONS}\label{DISCUSSION AND CONCLUSIONS}

The Galactic Astronomy and Galactic Dynamics communities are greatly looking forward to the second data release of the Gaia astrometric satellite. The data from this mission are expected to unveil  the phase space structure of our Galaxy, curing a sort of kinematic ``blindness'' we have had until now. Gaia will reveal the transverse motion dimensions of the phase-space of known stellar streams, and is expected to reveal more stream structures of low contrast that remain hidden in present-day star-counts surveys.

In this paper we presented a geometrical procedure that successfully gauges the Sun's velocity $\bmath{\textbf{V}}_{\odot}$ by allowing one to exploit a very basic behaviour of low-mass streams: that the proper motion of the stars should be closely directed along the structure in a frame that is at rest with respect to the Galaxy. Any perpendicular motion of the stream stars arises (primarily) due to the reflex motion of the observer. This effect is not correlated with Sun's Galactic distance $R_{\odot}$ value. The method was demonstrated using N-body simulated streams (degraded with Gaia-like uncertainties in proper motions and CFHT CFIS-like uncertainties in distance measurements). 
The reason for using low-mass streams for constraining the Solar motion using this method is simple. The high-mass (thick) streams like the Sagittarius stream formed from the disruption of dwarf galaxies, are highly dispersed in phase space \citep{2002MNRAS.332..915I, 2002ApJ...570..656J}. Although their broad trajectory could be curve fitted (or modelled using an orbit) for our purposes, the dispersion in the stream track would result in higher uncertainties in the measured Solar motion values.

Our method does not assume any Galactic mass model, which we view as a strength of the technique. If the Galactic potential were well known, it could clearly be used to refine the streams and hence obtain better constraints on $\bmath{\textbf{V}}_{\odot}$, but that would also require much more sophisticated modelling of individual tidal streams. 

It should also be noted that our analysis determines the Sun's velocity with respect to a sample of streams in the Galactic halo, and this velocity might turn out to be different from the velocity  measured with respect to the Galactic centre or with respect to the LSR for a variety of interesting astrophysical reasons. This could happen, if for instance, if Sgr $A^{\ast}$ is not at rest with respect to the Galaxy, or if the disk possesses significant non-circular motions, or if there is a bulk motion of the streams with respect to the disk (as might happen if there is a substantial ongoing accretion: e.g., the LMC or the Sgr dwarf). Using two independent measurement techniques might give us some insight about the relative motion between the dynamical centres of the inner Milky Way and the outer Milky Way (around which the streams actually orbit).

\section*{Acknowledgements}
The authors would like to thank the anonymous referee for very useful comments that contributed to the clarity and overall improvement of the paper.

\bibliographystyle{mnras}
\bibliography{main} % if your bibtex file is called example.bib

%%%%%%%%%%%%%%%%%%%%%%%%%%%%%%%%%%%%%%%%%%%%%%%%%%

% Don't change these lines
\bsp	% typesetting comment
\label{lastpage}
\end{document}